\documentclass[10pt,sigconf,letterpaper,nonacm]{acmart}

\PassOptionsToPackage{hyphens}{url}
\usepackage[hyphens]{url}
\usepackage{hyperref}
\hypersetup{breaklinks=true}
\usepackage{balance}

\usepackage{algorithm}
\usepackage{algorithmic}
\usepackage{amsfonts}
\usepackage{boxedminipage}
\usepackage{caption}
\usepackage{color}
\usepackage{colortbl}
\usepackage{enumerate}
\usepackage{epsfig}
\usepackage{graphicx}
\usepackage{listings}
\usepackage{multirow}
\usepackage{paralist}
\usepackage{pifont}
\usepackage{rotating}
\usepackage{tabularx}
\usepackage[usenames,dvipsnames]{xcolor}
\usepackage{xspace}
\usepackage{comment}
\usepackage{subfigure}
\usepackage{fancyvrb}

\usepackage{enumitem}

\usepackage{newtxtext,newtxmath}

\includecomment{notready}
\excludecomment{ready}
\begin{ready}
  \newcommand{\plonka}[1]{\typeout{\the\inputlineno: PLONKA: #1}}
\end{ready}
\begin{notready}
  \newcommand{\plonka}[1]{{\color{red}{\it plonka - #1}}\typeout{\the\inputlineno: PLONKA: #1}}
\end{notready}

\let\oldSim\sim
\renewcommand{\sim}{\raise.17ex\hbox{$\scriptstyle\oldSim$}}

\title{Is Protective DNS Blocking the Wild West?}

\author{David Plonka}
\email{plonka@wiscnet.net}
\affiliation{%
  \institution{WiscNet}
  \city{Madison}
  \state{Wisconsin}
  \country{USA}
}

\author{Branden Palacio}
\email{branden.palacio@marquette.edu}

\author{Debbie Perouli}
\email{despoina.perouli@marquette.edu}
\affiliation{%
  \institution{Marquette University}
  \city{Milwaukee}
  \state{Wisconsin}
  \country{USA}
}

\let\OldUrlFont\UrlFont \renewcommand{\UrlFont}{\small\OldUrlFont}

\begin{document}

\begin{abstract}

We perform a passive measurement study investigating how a Protective
DNS service might perform in a Research \& Education Network
serving hundreds of member institutions. Utilizing freely-available
DNS blocklists consisting of domain names deemed to be threats,
we test hundreds of millions of users' real DNS queries, observed
over a week's time, to find which answers would be blocked because
they involve domain names that are potential threats. We find the
blocklists disorderly regarding their names, goals, transparency,
and provenance, making them quite difficult to compare. Consequently,
these Protective DNS underpinnings lack organized oversight,
presenting challenges and risks in operation at scale.

\end{abstract}

\maketitle

\vspace{-0.8mm}
\section{Motivation \& Introduction}

Research \& Education Networks (RENs), worldwide, provide
Internet services that are either purpose-built or selected to
best meet the needs of their member institutions and users. Many
of these services implement security measures, {\em e.g.,}
global routing security~\cite{MANRS}, access control lists, and
firewalling. Likewise, a REN often provides core services such as
recursive Domain Name Service (DNS) resolvers, delivering correct
answers at high performance. Today, however, many enterprises and
institutions employ recursive DNS blocking, also known as Protective
DNS (PDNS), as a security measure. Such services selectively provide
a {\em wrong} answer (or refuse to answer) in response to a
query, having the goal of preventing a user from accessing Internet
hosts or sites deemed a threat to security or privacy.

The Internet Corporation for Assigned Names and Numbers
(ICANN) issued a report in May 2025~\cite{SAC127} that discusses
implementation strategies as well as consequences of DNS blocking
and PDNS. The report offers recommendations to entities implementing
blocking, organizations or governments, which include the entity
having a clear policy on what and how domains are blocked. ICANN's
report also recommends that users become aware that DNS blocking
is taking place through the use of DNS Extended Error codes by
operators, which would be a boon to troubleshooting and measurement.
Although the report makes a number of recommendations about the
operation of a recursive DNS blocking service, it does not comment on
the provenance nor evidence of disreputation of the sets of domain
names that it blocks. Thus we are motivated to measure those sets,
referred to as {\em blocklists}, and their performance on real
users' DNS queries.

Nikolich {\em et al.}~\cite{nikolich2025critical} explore the
relation of RENs to critical infrastructure. They observe that
RENs address Community Anchor Institutions network needs with
tailored services, including security, often in geographic areas
where commercial Internet Service Providers may have weak monetary
incentives. The paper argues for the importance of recognizing
RENs critical role through a new designation that would allow for
their continuous growth. We are further motivated to measure the
performance of PDNS as a candidate component of REN infrastructure.

U.S. federal policies already exist for services that sometimes
use PDNS in RENs. The Children's Internet Protection
Act (CIPA~\cite{CIPA}) requires schools and libraries, that receive
discounts from the federal E-Rate program, to have an Internet
safety policy that includes measures to block or filter access to
obscene, pornographic, and other content deemed harmful to minors.
Companies offer products~\cite{SecurlyCIPA,SecurlyGuestNetwork}
to help satisfy this need.  Secondly, under the purview of
U.S. Cybersecurity and Infrastructure Security Agency (CISA), the
non-profit that operates MS-ISAC~\cite{MSISAC} makes a Malicious
Domain Block and Reporting (MDBR) service available to U.S. state,
local, tribal, and territorial (SLTT) government agencies at no
cost to those entities~\cite{MDBR} and an additional option at
a cost~\cite{MDBRplus}.

\vspace{-0.9mm}
\section{Measurement}

Liu {\em et al.}~\cite{liu2024understanding} have identified
PDNS services worldwide and studied their behaviors via active
measurement.  Specifically, they synthesize and issue DNS queries to
reverse-engineer the services' behavior. Instead, we perform passive
measurements on real user traffic to determine which answers would be
blocked if a small selection of popular, freely-available blocklists
were employed: lists that ostensibly block only threats. That is,
it is not these lists' goal to block advertising, adult content,
gambling, etc.

{\bf Data Sources:}
We select three lists: $(1)$ Level Blue Labs' Open Threat
Exchange Phishing \& Scam domain names (OTX~\cite{OTX}),
the $(2)$ HaGeZi Threat Intelligence Feed (TIF~\cite{TIF}),
and the Université Toulouse Capitole's $(3)$ Prigent Malware
(Malware~\cite{Malware},~\cite{MalwareFTP},~\cite{MalwareGit}). These
DNS blocklists are maintained by subject matter experts
involving community input over time. The OTX list is the rare,
commercially-curated list that is open and free. Both the TIF and
Malware lists were selected based on each having a time series
archive ideal for study, while also suggesting discipline and
transparency in their curation. We collected versions of each list
dated June 13 or 15, {\em i.e.,} contemporaneous to the queries
described below.
\begin{figure}[h!]
\centering
\includegraphics[scale=0.34]{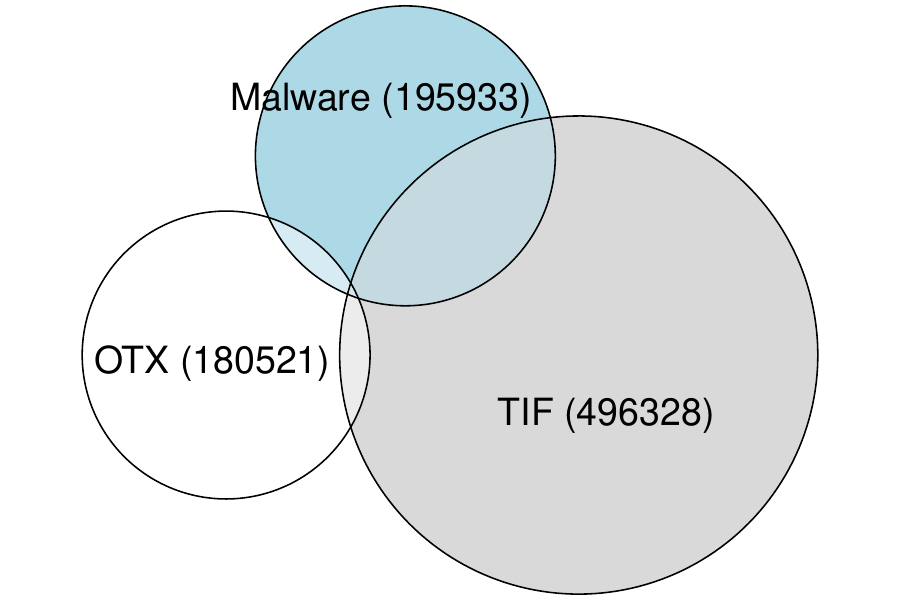}
\caption{Domain names in the blocklists, June 2025.} \label{fig:listsVenn} 
\end{figure}

Figure~\ref{fig:listsVenn} is a proportional Venn diagram summarizing the blocklists and their respective domain name counts. There is modest intersection of the sets: 2009 domain names occur in both OTX and TIF; 1670 domain names occur in both OTX and Malware; 55451 occur in both TIF and Malware; just 665 occur in all three sets.

We study a subset of REN users' DNS queries and their respective
answers in 7 contiguous days: June 19-25, 2025. This is done
by packet capture of the anycast servers' traffic containing
DNS query answers, over 890 million packets. The traffic
header and payload information were anonymized for user privacy.
Specifically, the client IP addresses were anonymized as well as select DNS query name
labels redacted, {\em e.g.,} where those labels might represent encoded IP
addresses as sometimes occurs in dynamic domain names. We then use the
treetop~\cite{Plonka08,treetop} tool, based on dnstop~\cite{dnstop},
to analyze and report DNS activity restricted to that involving
domain names in the blocklists, a further privacy measure.

\begin{figure}[h!]
\centering
\includegraphics[scale=0.34]{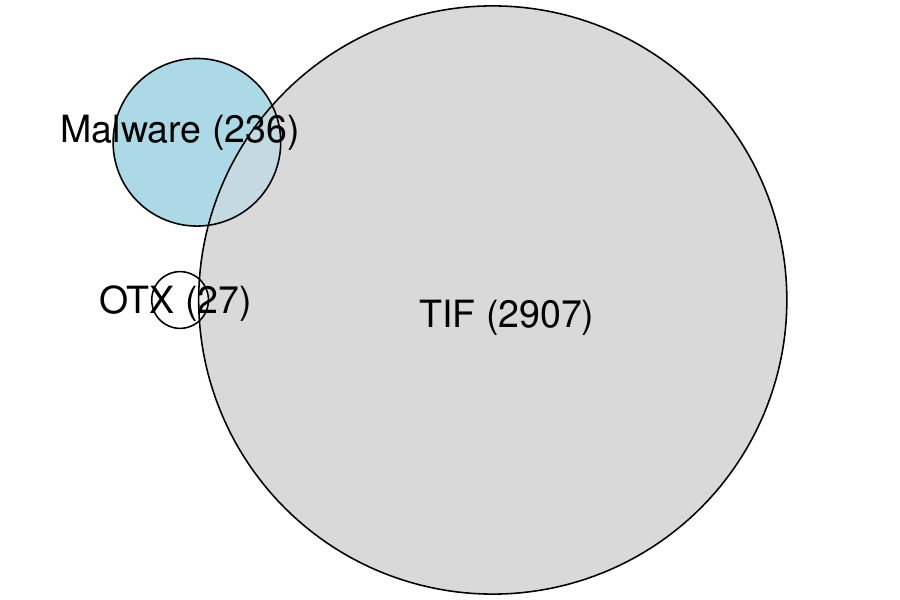}
\caption{Query names matching blocklists, June 2025.} \label{fig:resultsVenn} 
\end{figure}
{\bf Results:} Our results are briefly summarized in Figure~\ref{fig:resultsVenn}, a proportional Venn diagram of queried domain name counts matching each blocklist. In the 890 million response packets, we find 1.79 million unique query names: 27 of which matched OTX, 2907 matched TIF, and 236 matched Malware. Of those, only 3 query names were in both the OTX and TIF lists, while 27 were in both the TIF and Malware lists. There is no overlap in results between OTX and Malware lists.

\section{Discussion}

These early results suggest that popular freely-available blocklists,
pitched as alternatives to each other, are neither alike in content
nor in performance. This stems, in part, from their different
curation strategies and their curators' different meanings of terms
such as ``threat.'' The blocklists are diverse and sometimes at
odds with each other, {\em e.g.,} one list's curator blocks a
frequently-queried domain name in ongoing fashion, while another
list's curator ensures it is not blocked.

What is largely missing with these blocklists, as well as commercial
offerings, is {\em (a)} the ability to easily determine {\em why}
a domain name is present in a list, that we call {\em provenance,} and
{\em (b)} a common taxonomy
regarding goals of the blocking so that alternative implementations
can be evaluated, head to head. Performance evaluation is necessary
to make informed, timely decisions about filtered DNS operation,
either for individuals (who might choose another recursive DNS
resolver) or en masse, {\em e.g.,} for regional or state RENs that
serve myriad users. While the ICANN recommendations improve the
situation with respect to aiding measurement (prescribing that
blocking DNS services clearly indicate when they are refusing to
provide a correct answer), empirical performance evaluation of
Protective DNS blocking decisions remains a challenge.

Anticipating future commercial offerings, the introduction of
automated algorithms and/or AI-driven decisions about which domains to
block might further add to the confusion about what is blocked, why,
and even {\em when}, as automation drives frequent change. In
the same way that content filtering is selected and implemented for
schoolchildren (minors), the blocklists' purpose must be carefully
matched to the users and their roles and empower decision makers
to address problems when they arise while evaluating overall performance.

Looking forward, some RENs and their members, having
engineering staff with aligned goals, are likely well-positioned to
collectively develop and operate domain blocking privacy and security strategies
that best serve users at research and education institutions,
nationwide and worldwide. While such an endeavor clearly entails
significant effort, initially as well as in ongoing fashion,
the operational privacy and security benefits are commensurately
significant. Today, it simply is not clear that existing freely-available
open source blocklist-based solutions or commercial Protective
DNS products offer the high levels of privacy, responsible curation,
and transparency that RENs are trusted to provide to their members.

\newpage
\balance
\bibliographystyle{plain} 
\bibliography{main}

\appendix
\section{Ethics}

For privacy and security reasons, in this note we eschew identifying
specific domain names deemed threats by blocklists, whether or not
they were resolved in our measurement study. All client IP addresses
were anonymized or redacted in our collaborative analysis such that
query names cannot be associated with institutions or hosts.

\section{Acknowledgments}

Cameron Fronczak employed the OTX Phishing \& Scams blocklist in an
earlier study as part of his Master's thesis work.

Branden Palacio is supported through the CyberCorps Scholarship
for Service, National Science Foundation (NSF) award \#2235080.

This work was supported in part by the Open Science Grid (OSG)
consortium, which is supported by the National Science Foundation and
the U.S. Department of Energy Office of Science. Computational
resources were provided by the Open Science Pool (OSPool) of the OSG.

\end{document}